\documentclass{article}

\usepackage[utf8]{inputenc} 
\usepackage[T1]{fontenc}    
\usepackage{hyperref}       
\usepackage{url}            
\usepackage{booktabs}       
\usepackage{amsfonts}
\usepackage{amsmath}
\usepackage{amsthm}
\usepackage{tabularx}
\usepackage{float}
\usepackage{fullpage}
\usepackage{graphicx}
\usepackage{algorithm}
\usepackage{nicefrac}       
\usepackage{microtype}      
\usepackage{lipsum}
\usepackage[noend]{algpseudocode}
\usepackage[page]{appendix}

\newtheorem{lemma}{Lemma}[section]

\graphicspath{{./images/}}
\newcolumntype{A}{>{\raggedleft\arraybackslash}X}
\title{Sorting Lists with Equal Keys Using Mergesort in Linear Time}

\author{
  Albert Tedja
}
\date{}

\begin{document}
\maketitle

\begin{abstract}
This article introduces a new optimization method to improve mergesort's runtime complexity, when sorting sequences that have equal keys to $O(n log_2 k)$, where $k$ is the number of distinct keys in the sequence. When $k$ is constant, it is evident that mergesort is capable of achieving linear time by utilizing linked lists as its underlying data structure. Mergesort linked list implementations can be optimized by introducing a new mechanism to group elements with equal keys together, thus allowing merge algorithm to achieve linear time.
\end{abstract}

\section{Introduction}
The incremental nature of linked list allows only linear search and traversal, thus proving less useful compared to random-access arrays where they can benefit more from the locality of reference and $O(log_2{n})$ search algorithms such as binary or exponential search.  Nevertheless, linked lists $O(1)$ insertion and deletion can sometimes prove to be useful in some situations, in particular, mergesort.

Mergesort algorithm is usually implemented in two different ways. The most notable one is the top-down approach, in which the input sequence is split in half and each half is sorted independently and recursively, and finally merged to form the final sorted sequence. Top-down mergesort requires the length of the input sequence to be known in advance to optimally divide the sequence equally in half.  The second method is the bottom-up approach, where it iteratively creates sublists of length one and inserts them into a stack implementation. Sublists in stack are merged depending on certain heuristics.  This removes the requirement of knowing the length of the list in advance.  Panny and Prodinger \cite{Panny1995} provide a detailed analysis of top-down vs bottom-up mergesort and conclude that top-down mergesort provides a better performance although only by a slim margin.  Many linked list implementations keep track of an internal number of items in the list that it takes $O(1)$ to retrieve its length. However, due to the iterative nature of linked lists, the bottom-up approach of mergesort is chosen as the baseline of the optimization described here.

\section{Existing Optimizations and Implementations}

Natural mergesort is a variant of the bottom-up approach with optimizations to detect existing monotonic sublists in the original input sequence, also colloquially known as \textit{runs}.  A run is not split but rather kept and merged with its adjacent sublists. Brady \cite{Brady2005} introduces Runsort, that takes a step further by also identifying decreasing monotonic sublists and reverses their order before merging them.

Part of implementing an optimal natural mergesort algorithm is to determine the heuristic of when to merge sublists to maximize optimality in both space and time because each run has varying lengths. Shivers \cite{Shivers2002} proposes a solution by tracking the length of sublists and prioritizing merging shorter sublists.  Timsort \cite{Peters2002} approaches the problem in a similar fashion, but adds an additional layer by keeping a minimum threshold value that if a sublist's length is below a certain threshold, insertion sort is used before it is pushed into the stack. Knuth \cite[p.163]{Knuth1998} noted such optimization is worthwhile to reduce administrative overhead of short merges without affecting the asymptotic running time. These hybrid methods of using multiple sorting algorithms are quite common, and can be found in industrial-grade sorting libraries such as C++ STL's \texttt{std::sort} or some implementations of the GNU C's \texttt{qsort} based on the works of Bentley and Douglas \cite{Bentley1993}.

Wegner's trisort \cite{Wegner1982} attempts to sort linked lists with equal keys using the quicksort method with the average cost in the order of $O((m+n)log_2(\frac{n}{m}))$, where $n$ is the number of items occuring $m$ times. Similar to quicksort, however, it has the possibility to suffer from a bad pivot. It is especially even more likely to happen in linked lists as only either the $head$ or $tail$ can be chosen as the pivot.

\section{Baseline Algorithm}
Algorithm \ref{alg:stdbum} shown in Appendix \ref{appendix:alg} describes a straight-forward bottom-up mergesort algorithm implemented using linked lists without any optimization, such as detecting existing runs or comparing lengths of sublists. Unoptimized mergesort is chosen to demonstrate that the optimization described here is isolated and performs without the compounding effect of other optimizations. The algorithm hereafter shall serve as the baseline of the optimization presented in this article.

To summarize, the baseline algorithm treats each element as a sublist of length 1, and inserts sublists into a stack implementation. The algorithm does not maintain additional administrative computational complexities such as minimum runs in Timsort \cite{Peters2002} nor identifies shorter sublists in Shivers' \cite{Shivers2002} implementation, but rather resorts to a simpler approach similar to O’Keefe’s samsort \cite{OKeefe1982}

The process in which the stack determines if it needs to merge sublists is simply decided by the binary pattern of the stack counter. Let $c$ be the stack counter, and it counts the number of sublists inserted into the stack. The binary pattern of $c$, starting from the least significant digit, indicates the number of merges required before inserting the $(c+1)$-th sublist. For example, if $c = 7$, then $c = 0111_2$, then 3 merges are performed across three levels 1, 2, and 3, when inserting the 8th sublist. Merging is terminated once a 0 is encountered.  This solution is therefore opportunistic and performs merges as soon as possible, and allows minimal constant space requirement for up to $2^\alpha-1$ number of sublists for a stack with maximum depth of $\alpha$.

\section{Merging Using Linked Lists}
Using arrays as the underlying structure and buffer, mergesort's merging operation is known to be proportional to $m+n$ \cite[p.159]{Knuth1998}, where $m$ and $n$ are the lengths of the sublists to be merged. The disadvantage of this approach is the $O(n)$ space requirement for the buffer. Attempts to reduce the $O(n)$ space requirement, such as by Katajainen et al. \cite{Katajainen1996} or Huang et al. \cite{Huang1988}, increase the merging time complexity to $O(n log_2 n)$, which is not favorable.

Linked lists, however, naturally allow merging without an additional $O(n)$ space requirement.  This comes at a cost that linked list implementations do not normally benefit from the principle of locality, and searching for insertion points must be done linearly. Carlsson et al. \cite{Carlsson1993} introduce sublinear merging using a combination of arrays and linked lists by grouping blocks from input sequences $X$ and $Y$, where each block is computed by using binary and exponential searches alternatingly between $X$ and $Y$. The computed ranges of blocks are organized inside a linked list $Z$, such that traversing $Z$ outputs the merged sorted sequence, thereby preventing the need to copy or swap elements.

It is possible to merge two sublists with cost function proportional to $k$, where $k$ is the total number of distinct keys in the sublists, by using linked list as the underlying data structure. Instead of identifying monotonic ranges in input sequences, it identifies elements with equal keys. This reduces the overall mergesort time complexity to linear when sorting such particular datasets.

\section{Optimizing with Hop Pointers}
In addition to the ubiquitous linked list’s \textit{next} pointer, and its contrast \textit{previous} pointer in doubly linked lists, this article introduces a new concept called the \textit{hop} pointer. The hop pointer is remarkably simple as by default it simply points to itself, its host node. Given a sequence, and there exist elements with equal keys adjacent to each other, we define these contiguous set of elements with equal keys as a \textit{segment}. When such a segment exists in a sequence, then the hop pointer of the first node in the segment shall point to the last node in the segment.

\begin{figure}
\centering
\includegraphics[width=\textwidth]{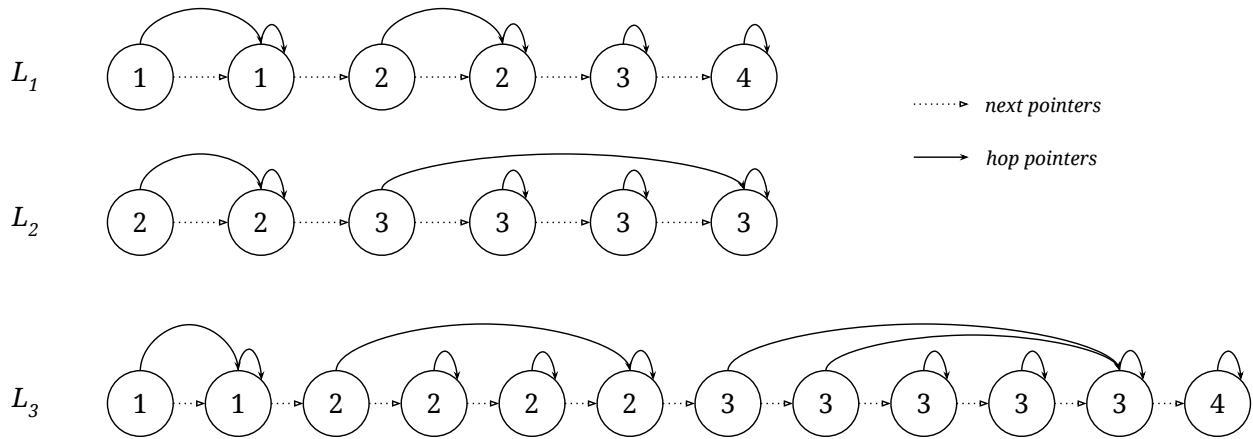}
\caption{Shows two lists $L_1$ and $L_2$ with hop pointers. $L_3 =$ {\footnotesize MERGE}$(L_1, L_2)$, from Algorithm \ref{alg:hopbum}. The second $3$ in $L_3$ is the first $3$ from $L_2$. Its hop pointer can be optionally set to itself after the merge, but not required for the optimization discussed here.}
\label{figure:hp}
\end{figure}

As observed in Figure \ref{figure:hp}, the hop pointer of the first element in a segment identifies the last node in the segment.  The purpose of hop pointers is to compress the list down to its distinct keys, allowing an iterative algorithm to \textit{skip forward} to the end of the segment, essentially treating an entire segment as one step.

\begin{lemma}
\label{mergepq}
Let $p$ and $q$ be the number of distinct keys in sorted lists $L_1$ and $L_2$, respectively. The cost of merging $L_1$ and $L_2$ using hop pointers is $O(p+q)$.
\end{lemma}

\begin{proof}
Hop pointers group elements with equal keys as one element. Consequently, the cost of iterating $L_1$ and $L_2$ is $O(p)$ and $O(q)$, respectively. An efficient merge algorithm requires iterating $L_1$ and $L_2$ each exactly once \cite[p.159]{Knuth1998}. The cost of merging is therefore $O(p+q)$.
\end{proof}

Merging two sequences no longer requires $O(|L_1|+|L_2|)$ but $O(p+q)$. Since $L_3 =$ {\footnotesize MERGE}$(L_1, L_2)$, we know that $k$, the number of distinct keys in $L_3$, cannot be greater than $p + q$, therefore $k \leq p+q$.

This optimization extends to subsequent merges thereafter, such as merging $L_3$ with $L_4$, $L_5$, ..., $L_n$. The complexity of merging is no longer dependent on the length of the lists but by the number of their distinct keys, thus allowing merge algorithm to perform more efficiently.

Implementing hop pointers on top of the baseline algorithm is quite straight-forward, and does not involve complex restructuring  of the code or logic rearrangement. Algorithm \ref{alg:hopbum} in Appendix \ref{appendix:alg} shows the optimized merge using hop pointers. A single hop pointer in every node is mostly sufficient for singly linked lists.  An additional \textit{hop-back} pointer can be optionally added if identifying a segment’s head node is necessary, for example, when appending one linked list to another, or in doubly linked lists where reverse traversals are permitted.

\section{Linear Time Complexity}

When $n = k$, every element is distinct, the conventional recurrence relation expression for mergesort $T(n) = n + n log_2 n$ is assumed. When $n > k$, i.e. there exist elements with equal keys, we define the recurrence relation to be $T(n) = 2T(\frac{n}{2}) + k$, because the cost of the merge function can be no more than $O(k)$. Therefore, the recurrence relation notation is:

$$T(n) = \begin{cases}
n + n log_2 n &n = k \\
2T\left(\frac{n}{2}\right) + k & n > k
\end{cases}$$

The base case $T(n) = 1$ for $n=1$ is omitted because it is already solved and implied in the $n = k$ relation, which becomes the new base case.

Solving the recurrence relation for $n > k$ by expanding $T\left(\frac{n}{2}\right)$ yields:

\begin{equation*}
\begin{split}
T(n) & = 2\left[2T\left(\frac{n}{4}\right)+k\right] + k \\
& = 4T\left(\frac{n}{4}\right) + 3k \\
& = 4\left[2T\left(\frac{n}{8}\right)+k\right] + 3k \\
& = 8T\left(\frac{n}{8}\right) + 7k \\
& = \dots \\
& = pT\left(\frac{n}{p}\right) + (p-1)k
\end{split}
\end{equation*}

When $\frac{n}{p} = k$, $p = \frac{n}{k}$:

\begin{equation*}
\begin{split}
T(n) & = \frac{n}{k}T(k)+(\frac{n}{k}-1)k \\
& = \frac{n}{k}(k + k log_2 k)+n-k \\
& = n + n log_2 k+n-k \\
& = 2n + n log_2 k - k \\
\end{split}
\end{equation*}

When $k=n$, i.e. every element is distinct, the relation above reduces back to the conventional mergesort recurrence relation $T(n) = n + n log_2 n$. However, if $k$ remains constant, the mergesort runtime complexity becomes $O(n log_2 k)$ or simply $O(n)$.

\section{Empirical Result}

The following tests are conducted to measure the number of comparisons between the baseline implementation and the \textit{hoptimized}\footnote{In the spirit of optimization, given a string "hop\textrm{-}optimized", omitting the redundancy of the dash delimiter from the English language, finding the longest repeated non-overlapping substring yields "op". The string can then be reduced to "hoptimized".} implementation.  There are 3 types of sample sets tested:

\begin{itemize}
\item \textit{Shuffled} - Shuffled sequences where every element is distinct.

\item \textit{Sawtooth} - Repeated sawtooth pattern with values up to $k$.

\item \textit{K-Distinct} - Shuffled version of \textit{Sawtooth}.
\end{itemize}

Each dataset is executed over 100 random permutations with input length $n$ exponentially increasing from $2^7$ to $2^{22}$.  The total numbers of comparison are recorded and presented in Table \ref{table:cmptotal}.

\begin{table}
\centering
\scriptsize
\begin{tabularx}{\textwidth}{cAAAAAA}
\toprule
$n$ & \multicolumn{2}{c}{Shuffled} & \multicolumn{2}{c}{Sawtooth} & \multicolumn{2}{c}{K-Distinct} \\
\cmidrule(lr){2-3} \cmidrule(lr){4-5} \cmidrule(lr){6-7}
 & Baseline & Hoptimized & Baseline & Hoptimized & Baseline & Hoptimized \\
\midrule
$2^7$ & 735 & 735 & 448 & 448 & 736 & 736 \\
$2^8$ & 1 725 & 1 725 & 1 024 & 1 024 & 1 727 & 1 727 \\
$2^9$ & 3 961 & 3 961 & 2 304 & 2 304 & 3 964 & 3 964  \\
$2^{10}$ & 8 946 & 8 946 & 5 120 & 5 120 & 8 948 & 8 948  \\
$2^{11}$ & 19 942 & 19 942 & 12 287 & 11 275 & 19 936 & 17 918 \\
$2^{12}$ & 43 974 & 43 974 & 28 668 & 23 556 & 43 969 & 36 068 \\
$2^{13}$ & 96 131 & 96 131 & 65 524 & 48 139 & 96 135 & 72 415 \\
$2^{14}$ & 208 673 & 208 673 & 147 424 & 97 306 & 208 640 & 145 153 \\
$2^{15}$ & 450 094 & 450 094 & 327 600 & 195 641 & 450 028 & 290 687 \\
$2^{16}$ & 720 704 & 720 704 & 720 704 & 392 312 & 965 586 & 581 725 \\
$2^{17}$ & 2 062 483 & 2 062 483 & 1 572 416 & 785 655 & 2 062 221 & 1 163 812 \\
$2^{18}$ & 4 387 094 & 4 387 094 & 3 406 848 & 1 572 342 & 4 386 381 & 2 328 143 \\
$2^{19}$ & 9 298 502 & 9 298 502 & 7 337 728 & 3 145 717 & 9 296 875 & 4 656 799 \\
$2^{20}$ & 19 645 532 & 19 645 532 & 15 723 520 & 6 292 468 & 19 641 712 & 9 314 026 \\
$2^{21}$ & 41 388 301 & 41 388 301 & 33 543 168 & 12 585 971 & 41 379 540 & 18 628 520 \\
$2^{22}$ & 86 971 029 & 86 971 029 & 71 278 592 & 25 172 978 & 86 950 900 & 37 257 365 \\
\bottomrule
\end{tabularx}
\caption{Total number of comparisons in Bottom-Up Mergesort: Baseline vs Hoptimized. $k=1024$}
\label{table:cmptotal}
\end{table}

\textit{Shuffled} sample sets do not show any improvement as shown in Table \ref{table:cmptotal}. This is expected as hop pointers do not offer any benefit when sorting sequences where each element is distinct. In \textit{Sawtooth} and \textit{K-Distinct} sample sets, there are noticeable improvements in that the number of comparisons slowly turns linear once $n$ grows larger than $k$. Dividing the total comparisons by $n$ yields a more interesting observation shown in Table \ref{table:cmppe}. The average number of comparisons per element slowly plateaus even as $n$ grows exponentially.

\begin{table}
\centering
\scriptsize
\begin{tabularx}{\textwidth}{cAAAA}
\toprule
$n$ & \multicolumn{2}{c}{Sawtooth} & \multicolumn{2}{c}{K-Distinct} \\
\cmidrule(lr){2-3} \cmidrule(lr){4-5}
 & Baseline & Hoptimized & Baseline & Hoptimized \\
\midrule
$2^7$ & 3.50000 & 3.50000 & 5.75000 & 5.75000 \\
$2^8$ & 4.00000 & 4.00000 & 6.74609 & 6.74609 \\
$2^9$ & 4.50000 & 4.50000 & 7.74219 & 7.74219 \\
$2^{10}$ & 5.00000 & 5.00000 & 8.73828 & 8.73828 \\
$2^{11}$ & 5.99951 & 5.50049 & 9.73438 & 8.74902 \\
$2^{12}$ & 6.99902 & 5.75098 & 10.73462 & 8.80566 \\
$2^{13}$ & 7.99854 & 5.87634 & 11.73523 & 8.83972 \\
$2^{14}$ & 8.99805 & 5.93909 & 12.73438 & 8.85944 \\
$2^{15}$ & 9.99756 & 5.97049 & 13.73376 & 8.87106 \\
$2^{16}$ & 10.99707 & 5.98621 & 14.73367 & 8.87918 \\
$2^{17}$ & 11.99658 & 5.99407 & 15.73350 & 8.87918 \\
$2^{18}$ & 12.99609 & 5.99801 & 16.73272 & 8.88116 \\
$2^{19}$ & 13.99561 & 5.99998 & 17.73238 & 8.88214 \\
$2^{20}$ & 14.99512 & 6.00097 & 18.73180 & 8.88255 \\
$2^{21}$ & 15.99463 & 6.00146 & 19.73130 & 8.88277 \\
$2^{22}$ & 16.99414 & 6.00171 & 20.73031 & 8.88285 \\
\bottomrule
\end{tabularx}
\caption{Average number of comparisons per element. Baseline vs Hoptimized. $k=1024$}
\label{table:cmppe}
\end{table}

\section{Conclusion}

Even though the implementation described here does not take advantage of the locality of reference or other optimizations, it is demonstrated that linear time mergesort is achievable.  By simply adding a $hop$ pointer in every node, mergesort can achieve linear time when sorting permutations of sequences with equal keys.

Many real-world datasets have equal keys, such as sorting cars by their makes or models, merchandise by their brands, or citizens by cities or countries. The stability of mergesort allows this optimization to reach an even greater purpose because it keeps the original ordering intact.

Hop pointers can be implemented in tandem with other sorting optimizations, such as detecting existing runs and better merging strategies, thus creating an even faster mergesort. Hop pointers, additionally, can be used to optimize other sorting algorithms such as insertion sort, or quickly determine the number of unique keys after sorting. They can also increase the speed of binary and exponential searches by further reducing the number of elements to be searched.

\bibliographystyle{alpha}  
\bibliography{main}

\begin{appendices}
\section{Algorithms}
\label{appendix:alg}

\begin{algorithm}
\caption{Baseline Bottom-Up Mergesort}
\label{alg:stdbum}
\begin{algorithmic}[1]
\Procedure{merge}{$a, b$} \Comment{Merge two lists $a$ and $b$}
	\If{$a = \oslash$} \textbf{return} $b$
	\ElsIf{$b = \oslash$} \textbf{return} $a$
	\EndIf
	\State $head \gets \oslash$
	\If{$a.value \leq b.value$}
		\State $head \gets a$
		\State $a \gets a.next$
	\Else
		\State $head \gets b$
		\State $b \gets b.next$
	\EndIf
	\State
	\State $p \gets head$
	\While{$a \neq \oslash \land b \neq \oslash$}
		\If{$a.value \leq b.value$}
			\State $p.next \gets a$
			\State $p \gets a$
			\State $a \gets a.next$		
		\Else
			\State $p.next \gets b$
			\State $p \gets b$
			\State $b \gets b.next$
		\EndIf
	\EndWhile
	\State
	\If{$a = \oslash$} $p.next \gets b$
	\ElsIf{$b = \oslash$} $p.next \gets a$
	\EndIf

	\State \textbf{return} $head$
\EndProcedure
\algstore{stdbum}
\end{algorithmic}
\end{algorithm}

\begin{algorithm}
\begin{algorithmic}[0]
\algrestore{stdbum}
\Procedure{mergesort}{$node$}
	\If{$node = \oslash \lor node.next = \oslash$} \textbf{return} $node$
	\EndIf
	\State $S \gets Stack$
	\State $c \gets 0$ \Comment{Stack counter}

	\While{$node \neq \oslash$}
		\State $next \gets node.next$
		\State $node.next \gets \oslash$
		\State $b \gets c$
		\While{$b \bmod 2 = 1$}
			\State $a \gets \Call{pop}{S}$
			\State $node \gets \Call{merge}{a, node}$
			\State $b \gets \lfloor b / 2 \rfloor$
		\EndWhile
		\State
		\State \Call{push}{S, node}
		\State $c \gets c + 1$
		\State $node \gets next$
	\EndWhile
	\State
	\If{$node = \oslash$} $node \gets \Call{pop}{S}$
	\EndIf
	\While{$S$ \textbf{not empty}}
		\State $a \gets \Call{pop}{S}$
		\State $node \gets \Call{merge}{a, node}$
	\EndWhile
	\State
	\State \textbf{return} $node$
\EndProcedure
\end{algorithmic}
\end{algorithm}

\begin{algorithm}
\caption{Optimized Bottom-Up Mergesort with Hop Pointers}
\label{alg:hopbum}
\begin{algorithmic}[1]
\Procedure{merge}{$a, b$}
	\If{$a = \oslash$} \textbf{return} $b$
	\ElsIf{$b = \oslash$} \textbf{return} $a$
	\EndIf

	\State $head \gets \oslash$
	\If{$a.value \leq b.value$}
		\State $head \gets a$
		\State $a \gets a.hop.next$
	\Else
		\State $head \gets b$
		\State $b \gets b.hop.next$
	\EndIf
	\State
	\State $p \gets head.hop$
	\While{$a \neq \oslash \land b \neq \oslash$}
		\If{$a.value < b.value$}
			\State $p.next \gets a$
			\State $p \gets a.hop$
			\State $a \gets a.hop.next$		
		\ElsIf{$a.value > b.value$}
			\State $p.next \gets b$
			\State $p \gets b.hop$
			\State $b \gets b.hop.next$
		\Else
			\State $p.next \gets a$
			\State $p \gets b.hop$
			\State $tmp \gets a.hop.next$
			\State $a.hop.next \gets b$
			\State $a.hop \gets b.hop$
			\State $a \gets tmp$
			\State $b \gets b.hop.next$
		\EndIf
	\EndWhile
	\State
	\If{$a = \oslash$} $p.next \gets b$
	\ElsIf{$b = \oslash$} $p.next \gets a$
	\EndIf

	\State \textbf{return} $head$
\EndProcedure
\end{algorithmic}
\end{algorithm}

\end{appendices}

\end{document}